\documentclass[sigconf]{acmart}

\AtBeginDocument{%
  \providecommand\BibTeX{{%
    \normalfont B\kern-0.5em{\scshape i\kern-0.25em b}\kern-0.8em\TeX}}}

\begin{document}

\title{What do we know about Computing Education in Africa?  \\  A Systematic Review of Computing Education Research Literature}
\author{Ismaila Temitayo Sanusi}
\email{ismaila.sanusi@uef.fi}
\orcid{0000-0002-5705-6684}

\authornotemark[1]
\affiliation{%
  \institution{School of Computing, University of Eastern Finland}
  \streetaddress{P.O.Box 111, 80101}
  \city{Joensuu}
  %\state{State}
  \postcode{80101}
  \country{Finland}
}
\author{Fitsum Gizachew Deriba}
\email{fgizache@uef.fi}
\orcid{0000-0002-0699-0274}
\authornotemark[1]
\affiliation{%
  \institution{School of Computing, University of Eastern Finland}
  \streetaddress{P.O.Box 111, 80101}
  \city{Joensuu}
  %\state{State}
  \postcode{80101}
  \country{Finland}
  }
\renewcommand{\shortauthors}{Ismaila Temitayo Sanusi \& Fitsum Gizachew Deriba}
% ``acmart''

\begin{abstract}
Noticeably, Africa is underrepresented in the computing education research (CER) community. However, there has been some effort from the researchers in the region to contribute to the growing need of computing for all. In order to understand the body of works that emerged from the global south region and their area of focus in computing education, we conducted a systematic review of the literature. This research investigates the prominent CER journals and conferences to discern the kind of research that has been published and how much contribution they have made to the growing field. Of the 68 selected studies, 45 papers were from South Africa. The prominent aspect of computing in the literature is programming, which accounts for 43\%. We identified open areas for research in the context and discussed the implication of our findings for the development of CER in Africa.
\end{abstract}





\maketitle
\begin{figure}
  \begin{minipage}[t]{0.52\textwidth}
    \centering
    \includegraphics[width=\linewidth]{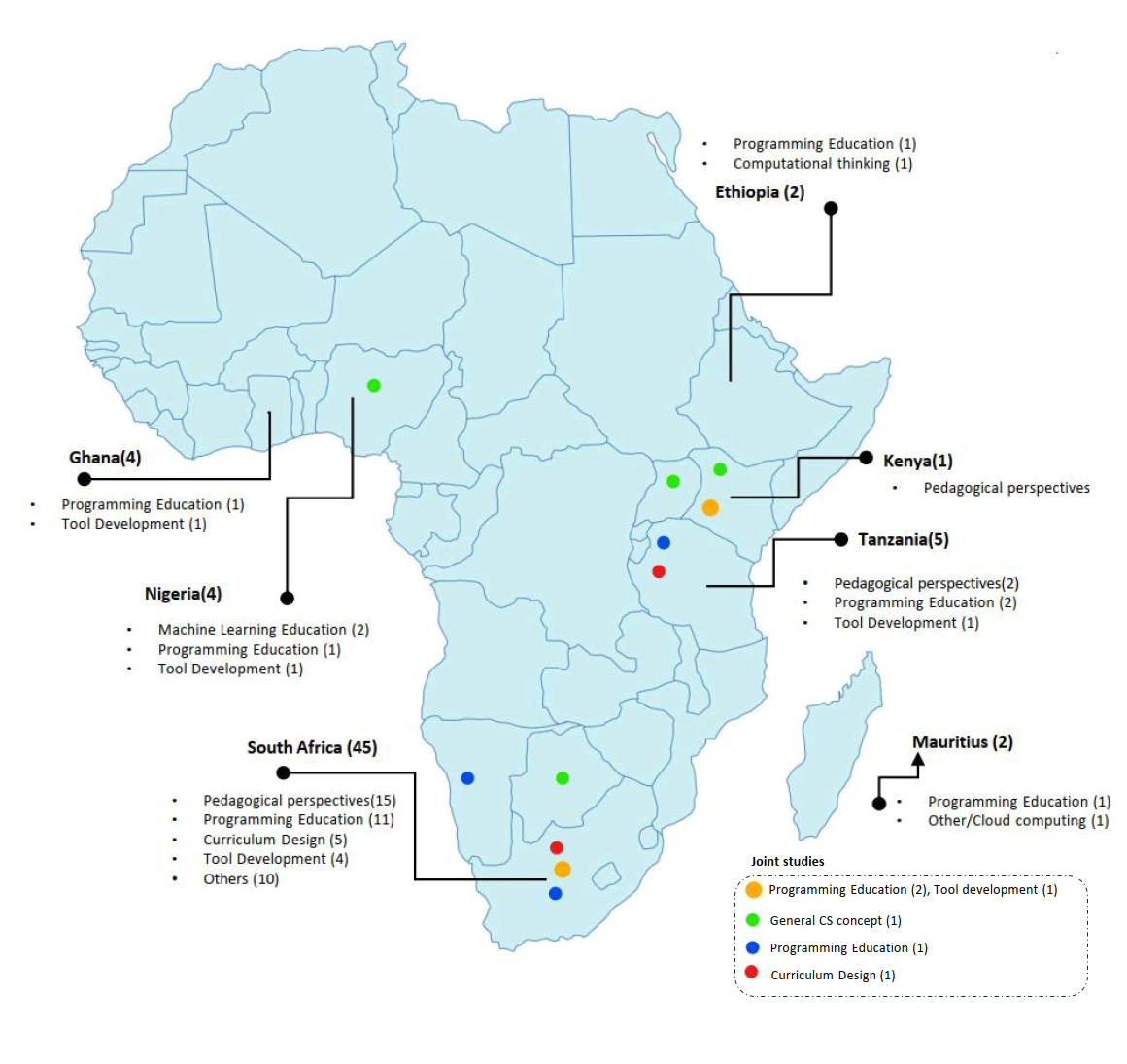}
    \caption{Landscape of Computing Education Research in Africa}
  \end{minipage}
\end{figure}

\section{Introduction}

Researchers claimed that the "CS for All movement is in full swing" \cite{payton2019reality}. Meanwhile, little evidence exist to show how the initiative is promoted in the world's second-largest and second-most populous continent, Africa. For over a decade, research within the computer science education (CSEd) community has decry poor geographic representation and the adverse effect it may have on computing education \cite{chvalovsky1978computer,wollowski2009expanding, becker2021expanding}. The underrepesentation of the region was also reported in a recent study by Brecker et al. \cite{becker2021expanding} who noted that one percent of ACM membership and less than one percent of SIGSCE membership are from Africa. While being a member of these professional bodies is not a prerequisite to publish in CSEd publication outlet, there is a report of significantly low participation and representation in computing education in Africa \cite{becker2021expanding}. 

Increasingly, broadening participation, inclusion, and diversity has become common terms in computing education. However, the African region remains underresearched and less represented in the discussion regarding computing education and its development across all levels from K-12 to higher education institutions (HEIs) \cite{agbo2023computing}. Addressing this obvious gap will contribute to high-quality research and equitable research that meets the teaching of all students computing. According to Jurado de Los Santos et al. \cite{jurado2020term} equity is related to the need established by many countries. As a result, it is imperative to explore learners with various needs and identify opportunities to support them within their contexts to participate in learning computing \cite{gizachew2022design}. To provide evidence for education policymakers and relevant stakeholders on how to achieve equitable access, participation, and learning experience in computing among all students, researchers has call for equitable enabling research \cite{mcgill2023conducting, fletcher2021cape}. In achieving equitable computing education, it is important to understand across different contexts and regions how learners, educators, education policymakers among other relevant stakeholders can be supported to successfully implement computing education. 

While it has been established that Africa has been greatly underrepresented in the global contribution to computing education \cite{mcgill2023meeting}, there are still some significant contributions to the field from the region. To understand the existing research that have been conducted on computing education research in Africa, we investigated major international computing education outlets. This paper provides insights from a systematic literature review, specifically on CER in Africa over the years, making at least the following contributions: 
\begin{itemize}
    \item Identifying the aspects of computing education that have been the focus of CER in Africa
    \item Summarising the computing education research practices within the African community
    %\item Highlighting the evidence that has been reported with regard to these developments
    \item Discussing the implication of the findings for development of CER in Africa
    \item Identifying how the results of this exposition will advance contributions of CER in Africa
    \item Discussing broadening of participation in computing education through the inclusion of understudied region to encapsulate holistically what equity-enabling research entails
    \item Research gaps in computing education in the African context.
\end{itemize}

Figure 1 presents the landscape of CER in Africa based on the result of our analysis.

%%%%%%%%%%%%%%%%%%%%%%%%%%%%%%%%%%%%%%%%%%%%%%%%%%%%%%%%%%%%%%%%%%%%%%%%%%%%%%%%%%%%%%%%%%
%%%%%%%%%%%%%%%%%%%%%%%%%%%%%%%%%%%%%%%%%%%%%%%%%%%%%%%%%%%%%%%%%%%%%%%%%%%%%%%%%%%%%%%%%%
%%%%%%%%%%%%%%%%%%%%%%%%%%%%%%%%%%%%%%%%%%%%%%%%%%%%%%%%%%%%%%%%%%%%%%%%%%%%%%%%%%%%%%%%%%

    \section{Methodology}
\label{sec:Method}

This study conducted a systematic literature review by following the guidelines proposed by Kitchenham \cite{kitchenham2004procedures}. As adopted by prior studies in computing education research \cite{luxton2018introductory}, this review, adopts a structured process which includes: Identifying research questions, Performing databases search, Selecting studies, Filtering the studies by evaluating their pertinence, Extracting data, Synthesizing the results and Writing the review report.

\subsection{Identifying research questions}

Based on the study aim, this study identified three research questions to guide the study. 
\begin{description}
\item [RQ1.] What are the characteristics of CER papers from Africa?
%\item [RQ2.] What research designs, data collection methods, and data analysis methods are applied to CER in Africa?
\item [RQ2.] What aspects of computing education have been the focus of literature from Africa?
\item [RQ3.] What open research areas need exploration in computing education in Africa?
\end{description}

\subsection{Performing databases search}
The literature search was performed in July 2023. Since this research focuses on understanding the research that emanates from Africa, this research adopts “Africa” as the sole search term within the whole document or field. The use of Africa as the only search term is justifiable since the databases searched focus only on computing education research. This study is not interested in specific fields of interest such as programming, or artificial intelligence (AI), etc. Hence, the authors of this research believe it is sufficient to proceed with a single search term. 

We used 10 different databases as shown in Table 1. These include the proceedings of the ACM Conference on Innovation and Technology in Computer Science Education (ITiCSE), ACM Conference on International Computing Education Research (ICER), ACM Technical Symposium on Computer Science Education(SIGCSE TS), Koli Calling International Conference on Computing Education Research (KOLI), WiPSCE Conference on Primary and Secondary Computing Education Research (WiPSCE), ACM Global Computing Education Conference (CompEd), Australasian Computing Education Conference (AUS-CE), Computer Science Education Research Conference (CSERC) and journal articles from ACM Transactions on Computing Education (TOCE) and Computer Science Education (CSE).  

CSERC is mostly hosted in Europe but has been hosted once in South Africa in 2016. Till date, none of the other listed conferences has been hosted in Africa. With CompEd created to be typically hosted outside North America or Europe, it may be hosted in Africa in the coming years. While other CSEd publication forums exist, the 10 outlets we consider in this study have been described as central forums in the CSEd field. Past research has utilized some of these publication forums \cite{lunn2021computer, hao2019systematic} to understand the development of computing education. 

\begin{table}[h!]
\centering
\captionof{table}{Database search}\label{tab:database}
\resizebox{.9\linewidth}{!}{%
\begin{tabular}{|l|l|l|l|l|} 
\hline
ID  & Publication outlet   & Search result    & Possible relevant results & Relevant papers   \\ 
\hline
1   & SIGCSE TS    & 357 & 15            & 9             \\ 
\hline
2   & ITiSCE    & 123 & 32             & 17           \\ 
\hline
3   & ICER    & 46 & 0             & 0           \\ 
\hline
4   & KOLI         & 27         & 12               & 10           \\ 
\hline
5 & WiPSCE & 11     & 2          & 2        \\
\hline
6   & CompED    & 3 & 0            & 0           \\ 
\hline
7   & CSE        & 61 & 6               & 6            \\ 
\hline
8 & ACM TOCE & 79     & 4             & 2         \\
\hline
9 & ACE [AUS-CE] & 8     & 0             & 0         \\
\hline
10 & CSERC & 27     & 26             & 22         \\
\hline
 & Total & 742     & 97             & 68         \\
\hline
\end{tabular}
}
\end{table}

\subsection{Selecting studies}
Based on the identified search terms and publication outlets, the first author searched and retrieved a total of 742 studies. Most of our searches returned articles that reported African-Americans as part of their participants. We did not consider a few studies that combined data from another continent outside Africa, such as \cite{buchele2013two} which utilized samples from Ghana and the USA. While we recognize that such collaboration broadens and encourages participation in the CSEd community as opined by \cite{becker2021expanding}, this study is interested in studies focusing on subjects within the African context. Future studies may explore studies that combine participants across continents or regions.

\subsection{Inclusion and exclusion criteria}
Before proceeding with the selection process for study inclusion, we established specific inclusion and exclusion criteria to identify potential papers.

\subsubsection{Inclusion criteria} All articles to be included should be primary studies conducted with samples or participants within the African continent and written in the English language.
Note that some papers consider different countries in Africa and they were included in our analysis (e.g. \cite{tshukudu2023investigating}). We included these papers because our interest is to understand works specifically focused on African participants. 
\subsubsection{Exclusion criteria}
Secondary studies, studies conducted in non-English languages, studies comparing Western and African populations, and studies that combine data from populations other than those in Africa are excluded from the study.
\subsection{Filtering the studies by evaluating their pertinence}
 After potential papers were identified (n=742), all articles were exported to Rayyan systematic tools for further identification of relevant studies \cite{johnson2018rayyan}. Two authors independently scanned the first 100 papers and categorized them as include, exclude, or maybe, by reading the abstract and title with the help of a filtering tool. The agreement between raters demonstrated high reliability according to Cohen's Kappa test ($\kappa = 0.89$), which results in near perfect according to \cite{belur2021interrater}. Any disagreements were addressed through discussion and resolution; in instances where the two authors held differing opinions on paper classification, they engaged in discussion until reaching a consensus. Finally, the author specifically selected papers that reported findings based on participants within the African context, which yielded 97 papers. 
 
\subsection{Full-text reading and Data extraction} 
After identifying potentially relevant papers, pertinent data was extracted into a spreadsheet by reading the entire document. Following, we perform coding to each study \cite{wubineh2023exploring}. This process involved retrieving details such as study objectives, study characteristics, and computing aspects as outlined in the selected articles. At this stage, 29 studies were further excluded because they did not meet the inclusion criteria; for example, studies that reported teaching computing without empirical evidence were not considered (e.g., \cite{anyanwu1978computer, bishop2004developing}).
Finally, 68 studies were included in this study, which are eligible for our analysis.

\subsection{Synthesizing the results} 
To synthesize the results of our review, we evaluate the individual findings of the 68 studies on spreadsheets and combine them based on their similar themes. Subsequently, we analyze and assess the reviewed studies to determine the outcomes of the research question \cite{pati2018write}. Finally, drawing upon the synthesized data, we generate insights from the findings and provide implications of the study.
\subsection{Writing the review report} 
Lastly, we present our findings based on the evidence derived from the synthesized results of CER included in the study. We succinctly summarize the CER findings within the African context, interpret their implications, and draw conclusions aimed at enriching the comprehension of the research landscape in Africa.

\section{Result}

\subsection{Characteristics of the Selected Papers}
\begin{figure}
  \begin{minipage}[t]{0.48\textwidth}
    \centering
    \includegraphics[width=\linewidth]{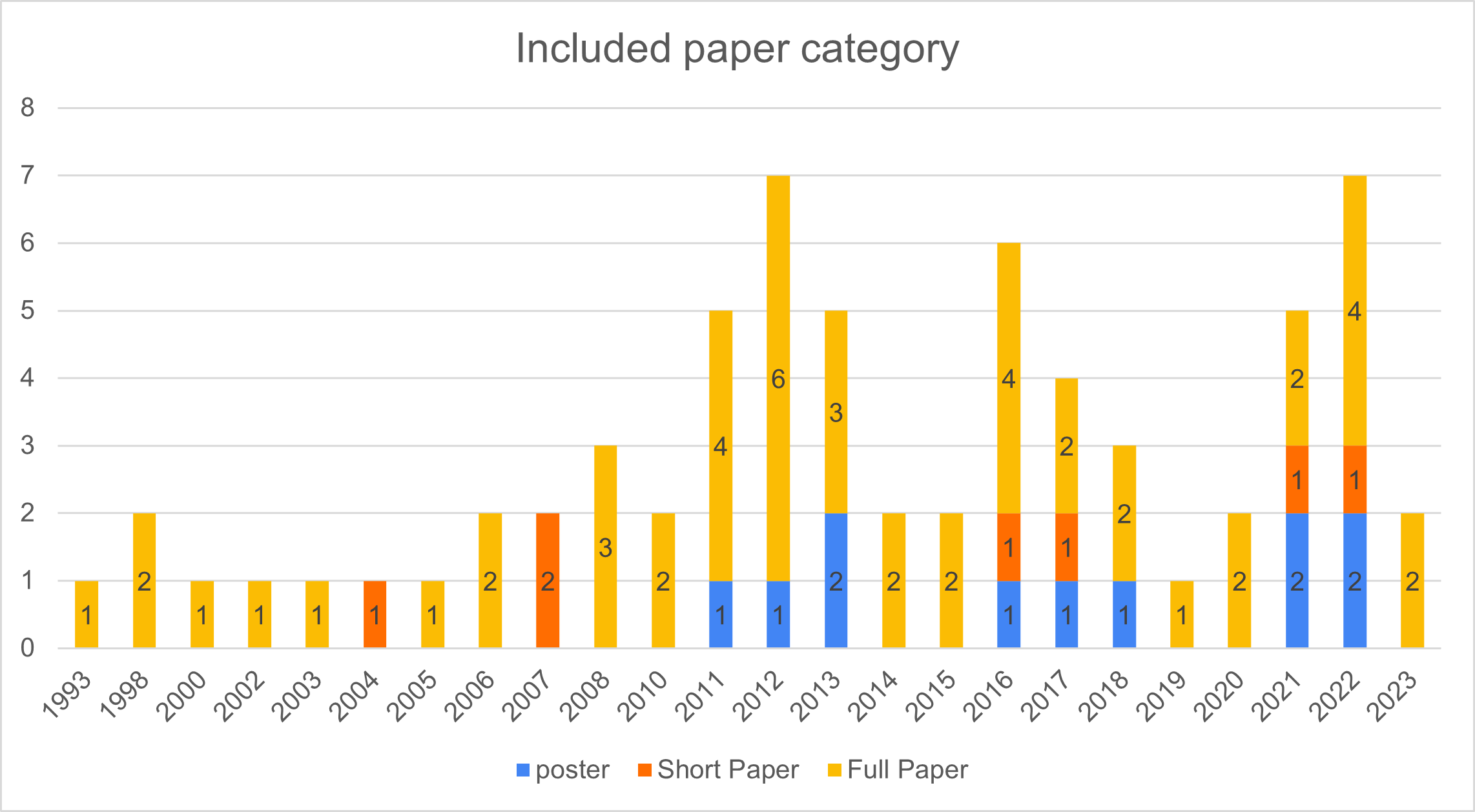}
    \caption{selected studies}
     \label{SS}
  \end{minipage}
\end{figure}
\subsubsection{Year of Publication and Submission categories}
As shown in Figure~\ref{SS}, the 68 papers we identified were published over 3 decades from 1993 -- 2023. The highest number of articles were published in 2012 and 2022 respectively with 7 papers. There appears to be increased submissions from 2012 even though they are not so significant. The total number of accepted papers per year in computing education publication outlets currently seems to be about 4 on average.  

With respect to the submission categories, most of the articles were published as full papers (n=50), followed by posters (n=11) and short papers (n=7). 

%\begin{table}[h!]
%\centering
%\captionof{table}{Submission categories}\label{tab:database}
%\resizebox{.7\linewidth}{!}{%
%\begin{tabular}{|l|l|l|l|l|} 
%\hline
%Categories   & XXXX    & XXXX & XXXX   \\ 
%\hline
%Full paper   & - & -            & -             \\ 
%\hline
%Short paper    & - & -             & -            \\ 
%\hline
%poster    & - & -            & -          \\ 
%\hline
%Panel         & -         & -               & -           \\ 
%\hline
%System and tools & -     & -            & -        \\
%\hline
%\end{tabular}
%}
%\end{table}

\subsubsection{Education Level and Context of Study Setting} As shown in Figure~\ref{MLL}, 76.5\% (n=52) of the studies were from Higher Education and 16.2\% (n=11) were from K-12 education out of which 8 studies focused on high school students. The remaining 7.5\% (n=5) of the studies did not specify the education level. 

Our analysis also shows that only 9 papers stated the geographic area (Urban vs Rural) where the study were conducted. Out of the 9 papers, 6 were in rural areas and 3 were conducted in rural and urban areas without comparing samples in the two distinct location. 

\subsubsection{Research Design}
\begin{figure}
  \begin{minipage}[t]{0.48\textwidth}
    \centering
    \includegraphics[width=\linewidth]{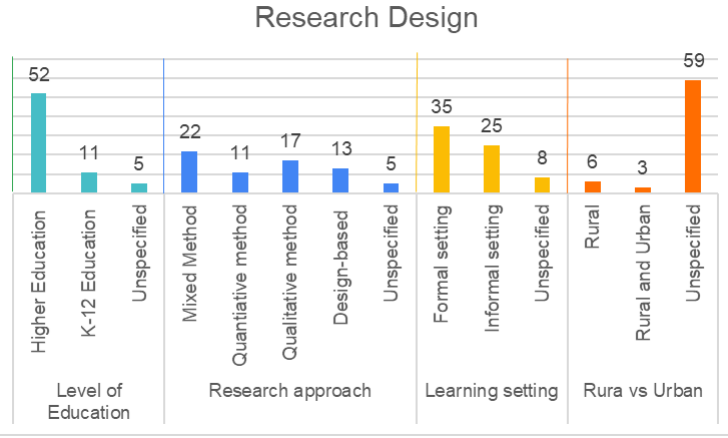}
    \caption{Research method, Education level and Learning setting}
     \label{MLL}
  \end{minipage}
\end{figure}

As illustrated in Figure~\ref{MLL}, this study indicate that the selected studies uses different research approaches. The majority (31\%; n=22) of the articles used mixed methods, followed by qualitative methods (25\%; n=17), design-based (19\%; n=13), and quantitative methods (16\%, n=11). The remaining 5 studies (7\%) do not explicitly state the approaches adopted.

    \subsection{Aspects of Computing Education}
\label{sec:ACE}
\begin{figure}
  \begin{minipage}[t]{0.50\textwidth}
    \centering
    \includegraphics[width=\linewidth]{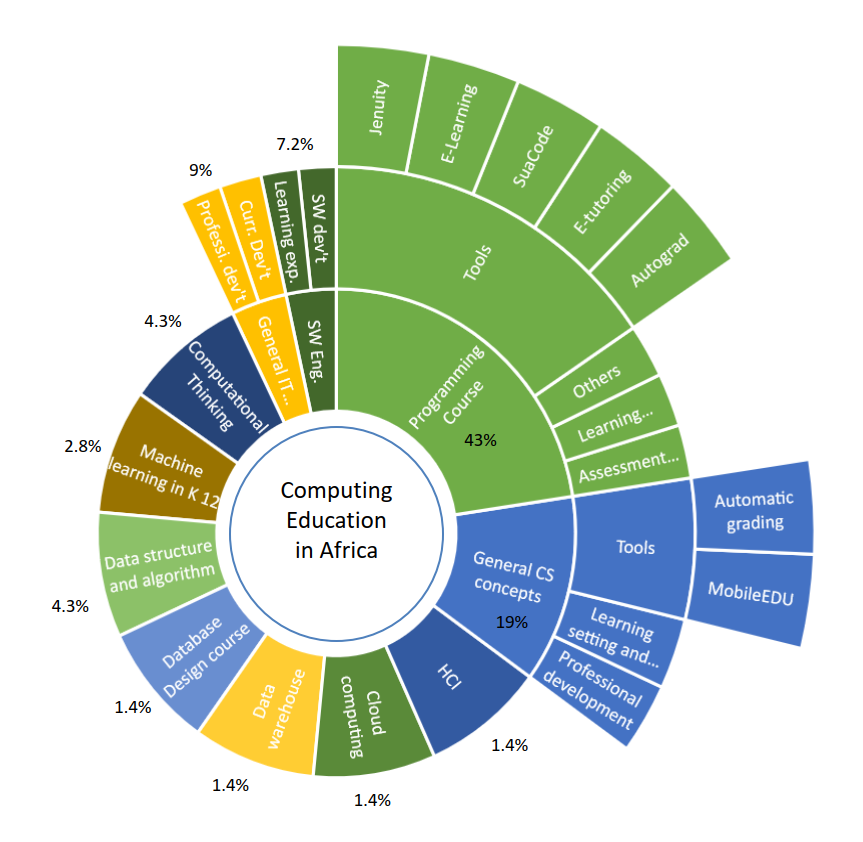}
    \caption{Aspects of Computing Education Research in Africa}
    \label{ACE}
  \end{minipage}
\end{figure}

As shown in Figure~\ref{ACE}, this study identified eleven aspects of computing education that have been the focus of literature from Africa. The eleven areas of focus are: programming, general CS concepts, software engineering, general IT concepts, computational thinking, machine learning for K-12, data structure and algorithm, database design course, data warehouse, cloud computing, and human-computer interaction (HCI).    

\subsubsection{Programming} Thirty-one studies (43\%) were programming-related. From them, seven studies focused on tool development such as E-tutoring \cite{horner2016tutoring}, E-learning \cite{messer2011use}, Saucode \cite{boateng2021suacode,boateng2019keep} , Autograd \cite{annor2021autograd}, Jenuity \cite{van2008jenuity}, and Recursive game generator \cite{anyango2021supporting}. Eight studies were on assessment and pedagogical strategies to programming \cite{chetty2015towards,pieterse2014managing,pieterse2017automatic,pieterse2013automated,apiola2012new,sanders2006mental,scholtz2010mental,gotschi2003mental}. Thirteen studies discussed different aspects and approaches to learning programming, including game-based learning \cite{oyelere2017integrating} and the use of blended learning \cite{bati2014blended}; experience, impact, and cooperativeness in pair programming \cite{benade2017pair,mentz2008effect,liebenberg2012pair}; effectiveness of scaffolding learning \cite{mbogo2016design}, the impact of iconic programming notation \cite{cilliers2005effect}, interactive learning in discussion forums \cite{pieterse2011student}, and describing of genetic programming \cite{pillay2004first}. The other category of programming-related discussed topics were student perception and experience \cite{govender2012students}, workshop experiences \cite{tshukudu2022broadening}, misconceptions and challenges in programming \cite{anyango2018teaching,pieterse2018triviality,motara2020obstacles}; the impact of parent technology adoption on students' education \cite{breedt2012student}.
\subsubsection{General CS Concepts} Thirteen studies (22.4\%) discussed general computer science concepts, including teaching tools (e.g., MobileEdu \cite{oyelere2016evaluating}, Automatic grading \cite{klein2011automated}), 4 studies on professional development such as teachers and students preconceptions \cite{tshukudu2023investigating,sanders2000fundamentals}, preparedness \cite{mbogo2019structured}, and capacity building \cite{apiola2015building}. Three studies discussed learning styles and settings, such as interactive learning \cite{kotze1998hypermedia}, informal learning \cite{klomsri2013social}, and learning styles \cite{galpin2007learning}. Another three studies related to curriculum development \cite{joseph2021ithinksmart, motara2020obstacles, pieterse2012participation}, and one study about student attitude on ethics \cite{boateng2019keep}.
\subsubsection{Software Engineering} Five studies (7.3\%) were related to software engineering courses: three discussed student involvement, aspiration, and teamwork in software development \cite{marshall2017student,liebenberg2016career, pieterse2012participation}; two on the impact of experiential learning \cite{marshall2016exploration}, and experiences \cite{pieterse2012intensive}  respectively.
\subsubsection{General IT Concepts} five studies (7.2\%) were focused on general IT-related concepts. From them, two \cite{tedre2008implementing,mentz2013empowering} on professional development; two \cite{laine2011refreshing,marais2016towards} on contextualizing and developing curriculum; one on perception \cite{govender2007challenges}.
\subsubsection{Computational Thinking (CT)} In this aspect, only three studies (4.4\%), focus on computational thinking (integration of ICT in curriculum \cite{kassa2022computational}, evaluation of framework \cite{gouws2013computational}, and education with Virtual reality \cite{joseph2021ithinksmart}). 
\subsubsection{Machine Learning for K-12}  Two studies on machine learning education focus on how to teach machine learning to young learners \cite{sanusi2021intercontinental, sanusi2023learning} within the K-12 levels. 
\subsubsection{Data Structure and Algorithm} Data structure and algorithm with three studies [4.4\%] in different concepts such as the influence of other languages and other subjects in algorithm courses \cite{rauchas2006language}, collaborative learning in algorithm courses \cite{nazir2019teaching}, and teaching approaches \cite{sanders2002teaching}. 
\subsubsection{Database Design Course} Database design course (on self-directed learning \cite{zyl2022developing}). The study aims to enhance SDL by identifying factors (e.g., considering various perspectives of peers when solving problems in cooperative learning groups) that emerged from the data. 
\subsubsection{Data warehouse} Data warehouse (on student view \cite{goede2016listening}). The author assessed the perceptions of students who have chosen a data warehousing 
module using critical systems heuristics,  to improve the module's design and 
ensure long-term benefits for all stakeholders.
\subsubsection{Cloud Computing} Cloud computing (student engagement \cite{nazir2021enhancing}). The study explored and implemented effective approaches for maintaining student engagement in an online and distance learning environment during the COVID-19 pandemic, focusing on the Cloud Platform Development coursework.
\subsubsection{Human-Computer Interaction (HCI)} HCI (theoretical knowledge on group project \cite{clarke1998teaching}). The study aimed to stress the value of hands-on experience with empirical methods in HCI 
education, enabling students to better apply theoretical knowledge to user-interface design.

\subsection{Open Research Areas for Computing Education Research}
\label{sec:ORACER}
Based on the RQ2 findings regarding the aspects of computing education researchers focused on, we identify some open areas of possible research for scholars in the context. 

\subsubsection{Emerging Technologies} Our findings indicate that research on understanding how students learn and educators teach emerging technologies such as artificial intelligence (AI) (including generative \& regenerative AI), machine learning (ML), and quantum computing is limited. As a result of these findings,  research focusing on teaching and learning these rapidly-emerging technologies needs to be fully explored including policies that supports their implementation \cite{temitayo2021teaching}. 

\subsubsection{Diversity Dimensions} According to McGill et al. \cite{mcgill2023conducting}, "diversity dimensions include various characteristics that have been historically used to differentiate groups, such as ethnicity, gender, religious beliefs, or socio-economic status." Africa is a continent with a very high linguistic diversity with an estimated 1500-2000 African languages \cite{Oneworld}. With these statistics, it is important to explore the complications that may arise based on language as past research has identified the links between language and learning computing among other diversity measures \cite{villegas2023effects}. Our findings showed that little or no report was given regarding the vulnerable groups, learning disability (e.g., neurodiversity), language complexity, and demographic influence in learning computing. 

\subsubsection{Ethics} Only a single paper focused on ethics in the retrieved articles. This finding is an indication that research on ethical implications has not been a subject of scrutiny. According to Sætra \& Danaher \cite{saetra2022each}, ethics plays a key role in the normative analysis of the impacts of technologies of different types, e.g., computer ethics, AI ethics, data ethics among others. Based on the proliferation of technology with specific impacts, ethical implications of these technologies should be understood by exploring several stakeholders in the region\cite{sanusi2022insight}.

\subsubsection{Educational Level} Our results suggest that majority of the studies were conducted in HEIs and only eleven (16.2\%) considered K-12 population. To this end, more research is required to explore various level and specific grade band from early childhood education through high school. Each educational stage such as primary, middle and high school have their own peculiarities which demands to be investigated to support the students within the cohort. Adult education is another domain to be examined. 

\subsubsection{Learning Context} Research has shown that learning context has implications for students learning of computing. For example, studies \cite{dunton2018examining, ma2023developing, sanusi2023machine} have shown that out-of-school programs are an essential tool for supporting and expanding K-12 computer science education. Hence, it is imperative to examine learning opportunities in-and-out of school system to support all learners within the developing region. 

\subsubsection{Use of Innovative Approaches} 
Pedagogy is not a static concept; rather, it evolves over time \cite{kunnari2019co}. There is a need to always rethink our pedagogy, particularly in a region with limited evidence of resources and evidence of effective pedagogical approaches that support students learning of computing. Researching into the use of novel approaches is in tandem with the global call for development of innovative ways to prepare learners with adequate knowledge and skills. For instance, in the ever-changing world, co-design provides an effective approach to foster engagement in learning\cite{sanusi2023preparing}, hence teachers must be willing to co-design with relevant stakeholders \cite{sanusi2022exploring}. Globally, there is a need for innovative ways to teach and introduce topics, learning materials, and ideas to learners regardless of educational level.  

\subsubsection{Teacher Education} Since computer science teacher preparation is critical to the development of CS education \cite{gal2010computer}, it is important to investigate this domain. With the continuous call for computing education for all learners, there should be increasing discussion about the educators who are to teach the learners. It is necessary to understand CS teacher education programs and professional learning opportunities available within the region.

\subsubsection{Conceptions of Computing} While studies \cite{xu2021exploring, grosse2019s} have investigated conceptions of various education stakeholders (e.g., students and educators) in relation to learning computing, there is a dearth of such research in the developing region of Africa. To this end, in relation to computing, extensive research that considers the perspectives of children, parents, education policymakers among other relevant stakeholders should be conducted to ensure the effective implementation of computing education initiatives.  

\section{Discussion}

\subsection{Insights generated from the findings}

%reflect based on study characteristics
The number of relevant studies we identified in our search of computing education publication outlets for our investigation supports the claim that there is a worrying lack of published work in international venues about CS education in Africa \cite{mcgill2023meeting}. The focus of the included papers on HEIs mostly suggests that less attention is paid to teaching and learning of CS education at k-12 levels. There is also a clear indication that geographical location of the participants is not considered important enough to be reported.

%aspect of computing
Regarding the aspect of computing that has been the focus of literature from Africa, we identified programming as the prominent topic. However, these studies on programming are HEI focused and none of them were conducted in the context of K-12. This finding shows that there is limited initiative on young learners and programming education which is consistent with existing research \cite{tshukudu2022broadening}. Other computing courses such as ML, CT and software engineering as shown in Subsection~\ref{sec:ACE} were barely researched. The limited research focus on ML education within the K-12 levels is consistent with reports in past research \cite{sanusi2021intercontinental}. However, no report of a study conducted on AI or ML in HEIs was found. 

%open areas
%The open areas highlighted in Subsection~\ref{sec:ORACER}, are critical to the development of CER in Africa.

This study provides an overview of the state of CSEd research in Africa. We specifically identified the aspects of computing education that have been the focus within the region. Our findings further provides insight into the areas of computing that needs exploration. These findings will contribute to broadening participation in computing. For instance, addressing the open areas identified in literature will contribute to meeting the needs of all learners to learn computing across educational level and learning settings. 

%\subsection{Implication of Study Findings}
%What is the study contribution
%why is this study important
%question bank
%raises the awareness within the community 

\subsection{Limitation and Future research}
While this research attempts to be exhaustive by investigating prominent computing education conferences and journals, Other computer science and education databases such as ACM, IEEE, and ERIC which may return more data were not considered. To broaden the search beyond "Africa," we recommend adding specific African country names and employing snowballing for enhanced study retrieval. Even though this limitation is recognized, our approach is not inadequate since most the retrieved studies mention the word "Africa" despite their focus on specific regions in Africa. 

To strengthen computing education in the UK and Ireland, UKICER was initiated. This is an indication that a CER conference hosted only in African countries be may be instituted. Future work could investigate computing education researchers in Africa and Where they publish. In addition, future research should explore why there is low research output in computing education research in Africa \cite{Sanusi2024exploring}.  

%While similar initiatives have begun in the region, more is needed. For example, an initiative to promote participation in Africa called CSEdBotswana was recently launched in Botswana \cite{tshukudu2023broadening}.

\section{Conclusion}

%\subsection{Closing Remarks}
This research employed a systematic review approach to understand the CER landscape in Africa with a focus on computing education literature. While it is recognized that some of the research conducted on computing education within the African context can be published in other forums (e.g. national or regional publication venues), this study holds the view that the CER-focused conferences and journal need to be investigated to get an overview of submissions from the continent. There could be a follow-up study to understand why conferences such as ICER has no record of contributions from Africa. While we identified panels have focused on enhancing computer and programming knowledge in developing countries including Africa \cite{khailany1982enhancing, sterling1978computer, adams2006approaches},  %papers that researchers of African descent and practice in Africa are being included on panels, 
there is no record of subject matters focusing specifically on CER in Africa. We did not also identify a report of Working Groups, organised by ITiSCE or CompEd focusing on CER in Africa.  %Future research may consider how much are African researchers or perspectives of scholars in the global south represented on the panels.
This paper fills a gap in the literature providing a picture of the state of the published literature about computing education research in Africa. 

\subsection{Some Guiding Questions for Future Research} Based on the eight open areas we identified for future research, researchers interested in the context may explore the following questions.  

\subsubsection{Emerging Technologies} How can we design models for young children's learning of programming?; How does generative AI supports learning of programming? \cite{deriba2023enhancing, aruleba2023integrating}; What opportunities exist for teaching quantum computing concepts in k-12 education in Africa?; How is cybersecurity education implemented in schools across the region of Africa?; How is AI curriculum implemented in schools in the developing regions of Africa? What are teachers practices and disposition and students learning experiences with respect to AI? How might we develop a framework for teaching generative AI \cite{thomas2024artificial}?

\subsubsection{Diversity Dimensions} What are the implications of language complexity for explaining computing concepts for CSEd in Africa?; How can learners with varying disabilities (including neurodiversity) types be supported to learn computing?; How much do we know about demographic influence in learning computing?; Are there ways to the address urban versus rural and public versus private schools dichotomy in computing learning?; What initiatives can instituted to assist street-connected children and youth in computing education?

\subsubsection{Ethics} What do children know about computer ethics (including emerging technology) ethics in Africa?; Is computer ethics education integrated into school curricula?; How do teachers perceive and teach computer ethics in schools?; What are the students' conceptions and attitudes toward computer ethics? 

\subsubsection{Educational Level} Does participating in K-12 computing courses have an impact on students’ future career choices in computing?; What do we know about learning computing skills in vocational schools and colleges\cite{Oladele2024What}? 

\subsubsection{Learning Context} What are the out-of-school initiatives that currently support computing education in Africa?; How can computing education be further promoted in schools? 

\subsubsection{Use of Innovative Approaches} Does co-design pedagogy support learning of computing?; How can contextualized resources be employed to support learning of computing ?; How much of support does unplugged activities (as a low-cost solution) foster for computing skills? 

\subsubsection{Teacher Education} Where is the computing education teacher program happening in Africa?; Why there are Low Research Output in Computing Education Research in Africa? \cite{Sanusi2024exploring}. 

\subsubsection{Conceptions of Computing} What are young children's conceptions of computing in African settings? \cite{oyedoyin2024young}. What do we know about the perspectives of various education stakeholders (e.g. parents, students, teachers, and policymakers) with respect to computing in Africa?

%computational thinking in compulsory education: State of play and practices from computing education

\bibliographystyle{ACM-Reference-Format}
\bibliography{mybib.bib}

\begin{thebibliography}{100}

\bibitem{adams2006approaches}
Elizabeth Adams, Doug Baldwin, Judith Bishop, John English, Pamela Lawhead, and Daniel Stevenson.
\newblock Approaches to teaching the programming languages course: A potpourri.
\newblock In {\em Proceedings of the 11th annual SIGCSE conference on Innovation and technology in computer science education}, pages 299--300, Bologna, Italy, 2006. ACM Press.

\bibitem{agbo2023computing}
Friday~Joseph Agbo, Maria Ntinda, Sonsoles L{\'o}pez-Pernas, Mohammed Saqr, and Mikko Apiola.
\newblock Computing education research in the global south.
\newblock In {\em Past, Present and Future of Computing Education Research: A Global Perspective}, pages 311--333. Springer, Willamette University, USA, 2023.

\bibitem{annor2021autograd}
Prince~Steven Annor, Edwin Kayang, Samuel Boateng, and George Boateng.
\newblock Autograd: Automated grading software for mobile game assignments in suacode courses.
\newblock In {\em Proceedings of the 10th Computer Science Education Research Conference}, pages 79--85, ACM New York, NY, USA, 2021. Association for Computing Machinery.

\bibitem{anyango2018teaching}
Jecton~Tocho Anyango and Hussein Suleman.
\newblock Teaching programming in kenya and south africa: What is difficult and is it universal?
\newblock In {\em Proceedings of the 18th Koli Calling International Conference on Computing Education Research}, pages 1--2, 2018.

\bibitem{anyango2021supporting}
Jecton~Tocho Anyango and Hussein Suleman.
\newblock Supporting cs1 instructors: Design and evaluation of a game generator.
\newblock In {\em Proceedings of the 26th ACM Conference on Innovation and Technology in Computer Science Education V. 1}, pages 115--121, ACM New York, NY, USA, 2021. Association for Computing Machinery.

\bibitem{anyanwu1978computer}
JA~Anyanwu.
\newblock Object-orientation in java for scientific programmers.
\newblock In {\em Papers of the SIGCSE/CSA technical symposium on Computer science education}, pages 37--40, 1978.

\bibitem{apiola2015building}
Mikko Apiola, Jarkko Suhonen, Abbi Nangawe, and Erkki Sutinen.
\newblock Building cs research capacity in sub-saharan africa by implementing a doctoral training program.
\newblock In {\em Proceedings of the 46th ACM Technical Symposium on Computer Science Education}, pages 633--638, 2015.

\bibitem{apiola2012new}
Mikko Apiola and Matti Tedre.
\newblock New perspectives on the pedagogy of programming in a developing country context.
\newblock {\em Computer Science Education}, 22(3):285--313, 2012.

\bibitem{aruleba2023integrating}
Kehinde Aruleba, Ismaila~Temitayo Sanusi, George Obaido, and Blessing Ogbuokiri.
\newblock Integrating chatgpt in a computer science course: Students perceptions and suggestions.
\newblock {\em arXiv preprint arXiv:2402.01640}, 2023.

\bibitem{bati2014blended}
Tesfaye~Bayu Bati, Helene Gelderblom, and Judy Van~Biljon.
\newblock A blended learning approach for teaching computer programming: design for large classes in sub-saharan africa.
\newblock {\em Computer Science Education}, 24(1):71--99, 2014.

\bibitem{becker2021expanding}
Brett~A Becker, Amber Settle, Andrew Luxton-Reilly, Briana~B Morrison, and Cary Laxer.
\newblock Expanding opportunities: Assessing and addressing geographic diversity at the sigcse technical symposium.
\newblock In {\em Proceedings of the 52nd ACM Technical Symposium on Computer Science Education}, pages 281--287, 2021.

\bibitem{belur2021interrater}
Jyoti Belur, Lisa Tompson, Amy Thornton, and Miranda Simon.
\newblock Interrater reliability in systematic review methodology: exploring variation in coder decision-making.
\newblock {\em Sociological methods \& research}, 50(2):837--865, 2021.

\bibitem{benade2017pair}
Trudie Benad{\'e} and Janet Liebenberg.
\newblock Pair programming as a learning method beyond the context of programming.
\newblock In {\em Proceedings of the 6th Computer Science Education Research Conference}, pages 48--55, 2017.

\bibitem{bishop2004developing}
Judith Bishop and Nigel Horspool.
\newblock Developing principles of gui programming using views.
\newblock In {\em Proceedings of the 35th SIGCSE technical symposium on Computer science education}, pages 373--377, 2004.

\bibitem{boateng2021suacode}
George Boateng, Prince~Steven Annor, and Victor Wumbor-Apin Kumbol.
\newblock Suacode africa: Teaching coding online to africans using smartphones.
\newblock In {\em Proceedings of the 10th Computer Science Education Research Conference}, pages 14--20, 2021.

\bibitem{boateng2019keep}
George Boateng, Victor Wumbor-Apin Kumbol, and Prince~Steven Annor.
\newblock Keep calm and code on your phone: A pilot of suacode, an online smartphone-based coding course.
\newblock In {\em Proceedings of the 8th Computer Science Education Research Conference}, pages 9--14, 2019.

\bibitem{breedt2012student}
Hugo Breedt and Vreda Pieterse.
\newblock Student confidence in using computers: the influence of parental adoption of technology.
\newblock In {\em Proceedings of Second Computer Science Education Research Conference}, pages 17--21, 2012.

\bibitem{buchele2013two}
Suzanne~Fox Buchele.
\newblock Two models of a cryptography and computer security class in a liberal arts context.
\newblock In {\em Proceeding of the 44th ACM technical symposium on Computer science education}, pages 543--548, 2013.

\bibitem{chetty2015towards}
Jacqui Chetty and Duan van~der Westhuizen.
\newblock Towards a pedagogical design for teaching novice programmers: design-based research as an empirical determinant for success.
\newblock In {\em Proceedings of the 15th Koli Calling Conference on Computing Education Research}, pages 5--12, 2015.

\bibitem{thomas2024artificial}
Thomas~KF Chiu, Zubair Ahmad, Murod Ismailov, and Ismaila~Temitayo Sanusi.
\newblock What are artificial intelligence literacy and competency? a comprehensive framework to support them.
\newblock {\em Computers and Education Open}, page 100171, 2024.

\bibitem{chvalovsky1978computer}
Vaclav Chvalovsky.
\newblock Computer science education at universities: the case of developing countries.
\newblock In {\em Papers of the SIGCSE/CSA technical symposium on Computer science education}, pages 41--47, 1978.

\bibitem{cilliers2005effect}
Charmain Cilliers, Andr{\'e} Calitz, and J{\'e}an Greyling.
\newblock The effect of integrating an iconic programming notation into cs1.
\newblock In {\em Proceedings of the 10th annual SIGCSE conference on Innovation and technology in computer science education}, pages 108--112, 2005.

\bibitem{clarke1998teaching}
Matthew~C Clarke.
\newblock Teaching the empirical approach to designing human-computer interaction via an experiential group project.
\newblock In {\em Proceedings of the twenty-ninth SIGCSE technical symposium on Computer science education}, pages 198--201, 1998.

\bibitem{deriba2023enhancing}
Fitsum~Gizachew Deriba, Ismaila~Temitayo Sanusi, and Amos~Oyelere Sunday.
\newblock Enhancing computer programming education using chatgpt-a mini review.
\newblock In {\em Proceedings of the 23rd Koli Calling International Conference on Computing Education Research}, pages 1--2, 2023.

\bibitem{dunton2018examining}
Sarah~T Dunton and Stephanie Rodriguez.
\newblock Examining the role of informal education in k-12 computing pathways \& cs education reform efforts.
\newblock In {\em Proceedings of the 49th ACM Technical Symposium on Computer Science Education}, pages 1064--1064, 2018.

\bibitem{fletcher2021cape}
Carol~L Fletcher and Jayce~R Warner.
\newblock Cape: A framework for assessing equity throughout the computer science education ecosystem.
\newblock {\em Communications of the ACM}, 64(2):23--25, 2021.

\bibitem{gal2010computer}
Judith Gal-Ezer and Chris Stephenson.
\newblock Computer science teacher preparation is critical.
\newblock {\em ACM Inroads}, 1(1):61--66, 2010.

\bibitem{galpin2007learning}
Vashti~C Galpin, Ian~D Sanders, and Pei-yu Chen.
\newblock Learning styles and personality types of computer science students at a south african university.
\newblock {\em ACM SIGCSE Bulletin}, 39(3):201--205, 2007.

\bibitem{gizachew2022design}
Fitsum Gizachew.
\newblock Design and implementation of amharic-based ide with high-level programming language.
\newblock {\em EAI Endorsed Transactions on Creative Technologies}, 9(30):e2--e2, 2022.

\bibitem{goede2016listening}
Roelien Goede.
\newblock Listening to the affected: Student views after starting a 4th year module in data warehousing.
\newblock In {\em Proceedings of the Computer Science Education Research Conference 2016}, pages 12--21, 2016.

\bibitem{gotschi2003mental}
Tina G{\"o}tschi, Ian Sanders, and Vashti Galpin.
\newblock Mental models of recursion.
\newblock In {\em Proceedings of the 34th SIGCSE technical symposium on Computer science education}, pages 346--350, 2003.

\bibitem{gouws2013computational}
Lindsey~Ann Gouws, Karen Bradshaw, and Peter Wentworth.
\newblock Computational thinking in educational activities: an evaluation of the educational game light-bot.
\newblock In {\em Proceedings of the 18th ACM conference on Innovation and technology in computer science education}, pages 10--15, 2013.

\bibitem{govender2012students}
Desmond~Wesley Govender and Irene Govender.
\newblock Are students learning object oriented programming in an object oriented programming course? student voices.
\newblock In {\em Proceedings of the 17th ACM annual conference on Innovation and technology in computer science education}, pages 395--395, 2012.

\bibitem{govender2007challenges}
Desmond~Wesley Govender and Manoj Maharaj.
\newblock Challenges with respect to the e-readiness of secondary school teachers in kwazulu-natal, south africa.
\newblock In {\em Proceedings of the 12th annual SIGCSE conference on Innovation and technology in computer science education}, pages 191--195, 2007.

\bibitem{grosse2019s}
Gregor Gro{\ss}e-B{\"o}lting, Yannick Schneider, and Andreas M{\"u}hling.
\newblock It's like computers speak a different language: Beginning students' conceptions of computer science.
\newblock In {\em Proceedings of the 19th Koli Calling International Conference on Computing Education Research}, pages 1--5, 2019.

\bibitem{hao2019systematic}
Qiang Hao, David~H Smith~IV, Naitra Iriumi, Michail Tsikerdekis, and Amy~J Ko.
\newblock A systematic investigation of replications in computing education research.
\newblock {\em ACM Transactions on Computing Education (TOCE)}, 19(4):1--18, 2019.

\bibitem{horner2016tutoring}
V~Horner and Patricia Gouws.
\newblock E-tutoring support for undergraduate students learning computer programming at the university of south africa.
\newblock In {\em Proceedings of the Computer Science Education Research Conference 2016}, pages 29--36, 2016.

\bibitem{johnson2018rayyan}
Nastasha Johnson and Margaret Phillips.
\newblock Rayyan for systematic reviews.
\newblock {\em Journal of Electronic Resources Librarianship}, 30(1):46--48, 2018.

\bibitem{joseph2021ithinksmart}
Friday Joseph~Agbo, Solomon Sunday~Oyelere, Jarkko Suhonen, and Markku Tukiainen.
\newblock ithinksmart: Immersive virtual reality mini games to facilitate students’ computational thinking skills.
\newblock In {\em Proceedings of the 21st Koli Calling International Conference on Computing Education Research}, pages 1--3, 2021.

\bibitem{jurado2020term}
Pedro Jurado~de Los~Santos, Antonio-Jos{\'e} Moreno-Guerrero, Jos{\'e}-Antonio Mar{\'\i}n-Mar{\'\i}n, and Rebeca Soler~Costa.
\newblock The term equity in education: A literature review with scientific mapping in web of science.
\newblock {\em International Journal of Environmental Research and Public Health}, 17(10):3526, 2020.

\bibitem{kassa2022computational}
Ermias~Abebe Kassa and Enguday~Ademe Mekonnen.
\newblock Computational thinking in the ethiopian secondary school ict curriculum.
\newblock {\em Computer Science Education}, 32(4):502--531, 2022.

\bibitem{khailany1982enhancing}
Asad Khailany.
\newblock Enhancing computer knowledge in less developed countries(panel discussion).
\newblock {\em ACM SIGCSE Bulletin}, 14(1):260, 1982.

\bibitem{kitchenham2004procedures}
Barbara Kitchenham.
\newblock Procedures for performing systematic reviews.
\newblock {\em Keele, UK, Keele University}, 33(2004):1--26, 2004.

\bibitem{klein2011automated}
Richard Klein, Angelo Kyrilov, and Mayya Tokman.
\newblock Automated assessment of short free-text responses in computer science using latent semantic analysis.
\newblock In {\em Proceedings of the 16th annual joint conference on Innovation and technology in computer science education}, pages 158--162, 2011.

\bibitem{klomsri2013social}
Tina Klomsri, Linn Greb{\"a}ck, and Matti Tedre.
\newblock Social media in everyday learning: How facebook supports informal learning among young adults in south africa.
\newblock In {\em Proceedings of the 13th Koli Calling International Conference on Computing Education Research}, pages 135--144, 2013.

\bibitem{kotze1998hypermedia}
Paula Kotz{\'e}.
\newblock Why the hypermedia model is inadequate for computer-based instruction.
\newblock {\em ACM SIGCSE Bulletin}, 30(3):148--152, 1998.

\bibitem{kunnari2019co}
Irma Kunnari, Jari Jussila, Vesa Tuomela, and Jukka Raitanen.
\newblock Co-creation pedagogy from cschool towards hamk design factory.
\newblock 2019.

\bibitem{laine2011refreshing}
Teemu~H Laine and Erkki Sutinen.
\newblock Refreshing contextualised it curriculum with a pervasive game project in tanzania.
\newblock In {\em Proceedings of the 11th Koli Calling International Conference on Computing Education Research}, pages 66--75, 2011.

\bibitem{liebenberg2012pair}
Janet Liebenberg, Elsa Mentz, and Betty Breed.
\newblock Pair programming and secondary school girls’ enjoyment of programming and the subject information technology (it).
\newblock {\em Computer Science Education}, 22(3):219--236, 2012.

\bibitem{liebenberg2016career}
Janet Liebenberg and Vreda Pieterse.
\newblock Career goals of software development professionals and software development students.
\newblock In {\em Proceedings of the Computer Science Education Research Conference 2016}, pages 22--28, 2016.

\bibitem{lunn2021computer}
Stephanie Lunn, Ma{\'\i}ra Marques~Samary, and Alan Peterfreund.
\newblock Where is computer science education research happening?
\newblock In {\em Proceedings of the 52nd ACM Technical Symposium on Computer Science Education}, pages 288--294, 2021.

\bibitem{luxton2018introductory}
Andrew Luxton-Reilly, Simon, Ibrahim Albluwi, Brett~A Becker, Michail Giannakos, Amruth~N Kumar, Linda Ott, James Paterson, Michael~James Scott, Judy Sheard, et~al.
\newblock Introductory programming: a systematic literature review.
\newblock In {\em Proceedings companion of the 23rd annual ACM conference on innovation and technology in computer science education}, pages 55--106, 2018.

\bibitem{ma2023developing}
Ruizhe Ma, Ismaila~Temitayo Sanusi, Vaishali Mahipal, Joseph~E Gonzales, and Fred~G Martin.
\newblock Developing machine learning algorithm literacy with novel plugged and unplugged approaches.
\newblock In {\em Proceedings of the 54th ACM Technical Symposium on Computer Science Education V. 1}, pages 298--304, 2023.

\bibitem{marais2016towards}
Craig Marais and Karen Bradshaw.
\newblock Towards a technical skills curriculum to supplement traditional computer science teaching.
\newblock In {\em Proceedings of the 2016 ACM Conference on Innovation and Technology in Computer Science Education}, pages 338--343, 2016.

\bibitem{marshall2017student}
Linda Marshall and Janet Liebenberg.
\newblock Student curriculum development buy-in: A study from an educational software development module.
\newblock In {\em Proceedings of the 6th Computer Science Education Research Conference}, pages 66--72, 2017.

\bibitem{marshall2016exploration}
Linda Marshall, Vreda Pieterse, Lisa Thompson, and Dina~M Venter.
\newblock Exploration of participation in student software engineering teams.
\newblock {\em ACM Transactions on Computing Education (TOCE)}, 16(2):1--38, 2016.

\bibitem{mbogo2019structured}
Chao Mbogo.
\newblock A structured mentorship model for computer science university students in kenya.
\newblock In {\em Proceedings of the 50th ACM Technical Symposium on Computer Science Education}, pages 1109--1115, 2019.

\bibitem{mbogo2016design}
Chao Mbogo, Edwin Blake, and Hussein Suleman.
\newblock Design and use of static scaffolding techniques to support java programming on a mobile phone.
\newblock In {\em Proceedings of the 2016 ACM Conference on Innovation and Technology in Computer Science Education}, pages 314--319, 2016.

\bibitem{mcgill2023meeting}
Monica~M McGill, Leigh~Ann DeLyser, Ismaila~Temitayo Sanusi, and Selina~Marianna Shah.
\newblock Meeting the needs of all learners through high quality k-12 computing education research.
\newblock In {\em Proceedings of the ACM Conference on Global Computing Education Vol 2}, pages 185--186, 2023.

\bibitem{mcgill2023conducting}
Monica~M McGill, Sarah Heckman, Christos Chytas, Michael Liut, Vera Kazakova, Ismaila~Temitayo Sanusi, Selina~Marianna Shah, and Claudia Szabo.
\newblock Conducting sound, equity-enabling computing education research.
\newblock In {\em Proceedings of the 2023 Working Group Reports on Innovation and Technology in Computer Science Education}, pages 30--56. 2023.

\bibitem{mentz2013empowering}
Elsa Mentz, Roxanne Bailey, Betty Breed, and Marietjie Havenga.
\newblock Empowering information technology teachers through professional development: an evaluation.
\newblock In {\em Proceedings of the 8th Workshop in Primary and Secondary Computing Education}, pages 37--38, 2013.

\bibitem{mentz2008effect}
Elsa Mentz, Johannes~L van~der Walt, and Leila Goosen.
\newblock The effect of incorporating cooperative learning principles in pair programming for student teachers.
\newblock {\em Computer science education}, 18(4):247--260, 2008.

\bibitem{messer2011use}
Orry~M Messer and Angelo Kyrilov.
\newblock The use of mediating artifacts in embedding problem solving processes in an e-learning environment.
\newblock In {\em Proceedings of the 16th annual joint conference on Innovation and technology in computer science education}, pages 390--390, 2011.

\bibitem{motara2020obstacles}
Yusuf~Moosa Motara.
\newblock Obstacles when teaching functional programming.
\newblock In {\em Proceedings of the 9th Computer Science Education Research Conference}, pages 1--2, 2020.

\bibitem{nazir2019teaching}
Sajid Nazir, Stephen Naicken, and James~H Paterson.
\newblock Teaching data structures through group based collaborative peer interactions.
\newblock In {\em Proceedings of the 8th Computer Science Education Research Conference}, pages 98--103, 2019.

\bibitem{nazir2021enhancing}
Sajid Nazir, Yovin Poorun, James Paterson, and Brian Hainey.
\newblock Enhancing student engagement through cloud computing coursework: Challenges and opportunities in the time of covid-19.
\newblock In {\em Proceedings of the 10th Computer Science Education Research Conference}, pages 110--112, 2021.

\bibitem{Oladele2024What}
Campbell~O. Oladele, Ismaila~Temitayo Sanusi, and Harrison~I. Atagana.
\newblock From palette to program: Exploring the impact of visual art backgrounds on programming achievement in college scratch classes.
\newblock In {\em Proceedings of the 2024 Innovation and Technology in Computer Science Education}. 2024.

\bibitem{Oneworld}
One World~Nations Online.
\newblock List of official, national and spoken languages of africa., 2023.

\bibitem{oyedoyin2024young}
Mayowa Oyedoyin, Ismaila~Temitayo Sanusi, and Musa~Adekunle Ayanwale.
\newblock Young children’s conceptions of computing in an african setting.
\newblock {\em Computer Science Education}, pages 1--36, 2024.

\bibitem{oyelere2016evaluating}
Solomon~S Oyelere, Jarkko Suhonen, Greg~M Wajiga, and Erkki Sutinen.
\newblock Evaluating mobileedu: third-year undergraduate computer science students' mobile learning achievements.
\newblock In {\em Proceedings of the 16th Koli Calling International Conference on Computing Education Research}, pages 176--177, 2016.

\bibitem{oyelere2017integrating}
Solomon~Sunday Oyelere, Jarkko Suhonen, and Teemu~H Laine.
\newblock Integrating parson's programming puzzles into a game-based mobile learning application.
\newblock In {\em Proceedings of the 17th Koli Calling International Conference on Computing Education Research}, pages 158--162, 2017.

\bibitem{pati2018write}
Debajyoti Pati and Lesa~N Lorusso.
\newblock How to write a systematic review of the literature.
\newblock {\em HERD: Health Environments Research \& Design Journal}, 11(1):15--30, 2018.

\bibitem{payton2019reality}
Jamie Payton, Jamika~D Burge, and Jill Denner.
\newblock The reality of inclusion: The role of relationships, identity, and academic culture in inclusive and equitable practices for broadening participation in computing education.
\newblock In {\em Proceedings of the 50th ACM Technical Symposium on Computer Science Education}, pages 494--495, 2019.

\bibitem{pieterse2013automated}
Vreda Pieterse.
\newblock Automated assessment of programming assignments.
\newblock {\em CSERC}, 13:4--5, 2013.

\bibitem{pieterse2017automatic}
Vreda Pieterse and Janet Liebenberg.
\newblock Automatic vs manual assessment of programming tasks.
\newblock In {\em Proceedings of the 17th Koli Calling International Conference on Computing Education Research}, pages 193--194, 2017.

\bibitem{pieterse2014managing}
Vreda Pieterse and Christoph Stallmann.
\newblock Managing a large tertiary computer science class.
\newblock In {\em Proceedings of the Computer Science Education Research Conference}, pages 79--90, 2014.

\bibitem{pieterse2018triviality}
Vreda Pieterse and Estelle Taylor.
\newblock On the triviality of the assignment statement.
\newblock In {\em Proceedings of the 7th Computer Science Education Research Conference}, pages 51--57, 2018.

\bibitem{pieterse2012intensive}
Vreda Pieterse, Lisa Thompson, Linda Marshall, and Dina~M Venter.
\newblock An intensive software engineering learning experience.
\newblock In {\em Proceedings of Second Computer Science Education Research Conference}, pages 47--54, 2012.

\bibitem{pieterse2012participation}
Vreda Pieterse, Lisa Thompson, Linda Marshall, and Dina~M Venter.
\newblock Participation patterns in student teams.
\newblock In {\em Proceedings of the 43rd ACM technical symposium on Computer Science Education}, pages 265--270, 2012.

\bibitem{pieterse2011student}
Vreda Pieterse and Isabel~J Van~Rooyen.
\newblock Student discussion forums: what is in it for them?
\newblock In {\em Computer Science Education Research Conference}, pages 59--70, 2011.

\bibitem{pillay2004first}
Nelishia Pillay.
\newblock A first course in genetic programming.
\newblock {\em ACM SIGCSE Bulletin}, 36(4):93--96, 2004.

\bibitem{rauchas2006language}
Sarah Rauchas, Benjamin Rosman, George Konidaris, and Ian Sanders.
\newblock Language performance at high school and success in first year computer science.
\newblock {\em ACM SIGCSE Bulletin}, 38(1):398--402, 2006.

\bibitem{saetra2022each}
Henrik~Skaug S{\ae}tra and John Danaher.
\newblock To each technology its own ethics: The problem of ethical proliferation.
\newblock {\em Philosophy \& Technology}, 35(4):93, 2022.

\bibitem{sanders2002teaching}
Ian Sanders.
\newblock Teaching empirical analysis of algorithms.
\newblock {\em ACM SIGCSE Bulletin}, 34(1):321--325, 2002.

\bibitem{sanders2006mental}
Ian Sanders, Vashti Galpin, and Tina G{\"o}tschi.
\newblock Mental models of recursion revisited.
\newblock In {\em Proceedings of the 11th annual sigcse conference on innovation and technology in computer science education}, pages 138--142, 2006.

\bibitem{sanders2000fundamentals}
Ian Sanders and Conrad Mueller.
\newblock A fundamentals-based curriculum for first year computer science.
\newblock {\em ACM SIGCSE Bulletin}, 32(1):227--231, 2000.

\bibitem{sanusi2021intercontinental}
Ismaila~Temitayo Sanusi.
\newblock Intercontinental evidence on learners’ differentials in sense-making of machine learning in schools.
\newblock In {\em Proceedings of the 21st Koli Calling International Conference on Computing Education Research}, pages 1--2, 2021.

\bibitem{sanusi2023machine}
Ismaila~Temitayo Sanusi.
\newblock {\em Machine Learning Education in the K--12 Context}.
\newblock PhD thesis, It{\"a}-Suomen yliopisto, 2023.

\bibitem{sanusi2022insight}
Ismaila~Temitayo Sanusi and Sunday~Adewale Olaleye.
\newblock An insight into cultural competence and ethics in k-12 artificial intelligence education.
\newblock In {\em 2022 IEEE global engineering education conference (EDUCON)}, pages 790--794. IEEE, 2022.

\bibitem{sanusi2023preparing}
Ismaila~Temitayo Sanusi, Joseph~Olamide Omidiora, Solomon~Sunday Oyelere, Henriikka Vartiainen, Jarkko Suhonen, and Markku Tukiainen.
\newblock Preparing middle schoolers for a machine learning--enabled future through design-oriented pedagogy.
\newblock {\em IEEE Access}, 2023.

\bibitem{sanusi2022exploring}
Ismaila~Temitayo Sanusi, Solomon~Sunday Oyelere, and Joseph~Olamide Omidiora.
\newblock Exploring teachers' preconceptions of teaching machine learning in high school: A preliminary insight from africa.
\newblock {\em Computers and Education Open}, 3:100072, 2022.

\bibitem{sanusi2023learning}
Ismaila~Temitayo Sanusi, Kissinger Sunday, Solomon~Sunday Oyelere, Jarkko Suhonen, Henriikka Vartiainen, and Markku Tukiainen.
\newblock Learning machine learning with young children: exploring informal settings in an african context.
\newblock {\em Computer Science Education}, pages 1--32, 2023.

\bibitem{Sanusi2024exploring}
Ismaila~Temitayo Sanusi and Ethel Tshukudu.
\newblock Exploring barriers and strategies to boost scientific output in computing education in africa: Early insights.
\newblock In {\em Proceedings of the 2024 Innovation and Technology in Computer Science Education}. 2024.

\bibitem{scholtz2010mental}
Tamarisk~Lurlyn Scholtz and Ian Sanders.
\newblock Mental models of recursion: investigating students' understanding of recursion.
\newblock In {\em Proceedings of the fifteenth annual conference on Innovation and technology in computer science education}, pages 103--107, 2010.

\bibitem{sterling1978computer}
Theodor Sterling, FK~Allotey, Asad Khailany, Maria Lucia~Blanck Lisboa, Ignacio Mijares, and Edward Robertson.
\newblock Computer science education in developing countries.
\newblock In {\em Papers of the SIGCSE/CSA technical symposium on Computer science education}, pages 179--179, 1978.

\bibitem{tedre2008implementing}
Matti Tedre, Fredrick~D Ngumbuke, Nicholas Bangu, and Erkki Sutinen.
\newblock Implementing a contextualized it curriculum: Ambitions and ambiguities.
\newblock In {\em Proceedings of the 8th International Conference on Computing Education Research}, pages 51--61, 2008.

\bibitem{temitayo2021teaching}
Ismaila Temitayo~Sanusi.
\newblock Teaching machine learning in k-12 education.
\newblock In {\em Proceedings of the 17th ACM conference on international computing education research}, pages 395--397, 2021.

\bibitem{tshukudu2022broadening}
Ethel Tshukudu, Sofiat Olaosebikan, Kenechi Omeke, Alexandrina Pancheva, Stephen McQuistin, Lydia~John Jilantikiri, and Maha Al-Anqoudi.
\newblock Broadening participation in computing: Experiences of an online programming workshop for african students.
\newblock In {\em Proceedings of the 27th ACM Conference on on Innovation and Technology in Computer Science Education Vol. 1}, pages 393--399, 2022.

\bibitem{tshukudu2023investigating}
Ethel Tshukudu, Sue Sentance, Oluwatoyin Adelakun-Adeyemo, Brenda Nyaringita, Keith Quille, and Ziling Zhong.
\newblock Investigating k-12 computing education in four african countries (botswana, kenya, nigeria, and uganda).
\newblock {\em ACM Transactions on Computing Education}, 23(1):1--29, 2023.

\bibitem{van2008jenuity}
Martin van Tonder, Kevin Naude, and Charmain Cilliers.
\newblock Jenuity: a lightweight development environment for intermediate level programming courses.
\newblock {\em ACM SIGCSE Bulletin}, 40(3):58--62, 2008.

\bibitem{villegas2023effects}
Ismael Villegas~Molina, Adrian Salguero, Shera Zhong, and Adalbert~Gerald Soosai~Raj.
\newblock The effects of spanish-english bilingual instruction in a cs0 course for high school students.
\newblock In {\em Proceedings of the 2023 Conference on Innovation and Technology in Computer Science Education V. 1}, pages 75--81, 2023.

\bibitem{wollowski2009expanding}
Michael Wollowski, M~Daniels, and {\AA}~Cajander.
\newblock Expanding opportunities: Assessing and addressing geographic diversity at the sigcse technical symposium.
\newblock 2009.

\bibitem{wubineh2023exploring}
Betelhem~Zewdu Wubineh, Fitsum~Gizachew Deriba, and Michael~Melese Woldeyohannis.
\newblock Exploring the opportunities and challenges of implementing artificial intelligence in healthcare: A systematic literature review.
\newblock In {\em Urologic Oncology: Seminars and Original Investigations}. Elsevier, 2023.

\bibitem{xu2021exploring}
Zhen Xu, Albert~D Ritzhaupt, Karthikeyan Umapathy, Yang Ning, and Chin-Chung Tsai.
\newblock Exploring college students’ conceptions of learning computer science: A draw-a-picture technique study.
\newblock {\em Computer Science Education}, 31(1):60--82, 2021.

\bibitem{zyl2022developing}
Sukie~Van Zyl and Elsa Mentz.
\newblock Developing deeper self-directed learning in database design: Factors that influence knowledge transfer.
\newblock In {\em Proceedings of the 17th Workshop in Primary and Secondary Computing Education}, pages 1--2, 2022.

\end{thebibliography}

%\appendix

\end{document}